\documentclass[longnamesfirst,apjl]{emulateapj}
\usepackage{apjfonts,natbib}
\tighten
\newcommand{\beq}{\begin{equation}}
\newcommand{\eeq}{\end{equation}}
\newcommand{\bea}{\begin{eqnarray}}
\newcommand{\eea}{\end{eqnarray}}

\begin{document}
\title{The Afterglow of Massive Black Hole Coalescence}
\author{Milo\v s Milosavljevi\'c$^1$ and E.~S.~Phinney$^1$}
\affil{$^1$Theoretical Astrophysics, California Institute of Technology, Mail Code 130-33, 1200 East California Boulevard, Pasadena, CA 91125.}
\righthead{THE AFTERGLOW OF BLACK HOLE COALESCENCE}
\lefthead{MILOSAVLJEVI\'C \& PHINNEY}
\begin{abstract}

The final merger of a pair of massive black holes in a galactic nucleus is 
compelled by gravitational radiation. 
Gravitational waves from the mergers of black holes of 
masses $(10^5-10^7)(1+z)^{-1}M_\odot$ at redshifts
of $1-20$ will be readily detectable by the 
{\it Laser Interferometer Space Antenna}, but an electromagnetic
afterglow would be helpful in pinpointing the source and its redshift.
Long before the merger, the binary
``hollows out'' any surrounding gas and shrinks slowly
compared to the viscous timescale of
a circumbinary disk. The inner gas disk is truncated at the radius where
gravitational torque from the binary balances the viscous torque, and 
accretion onto the black holes is diminished.  Initially, the inner
truncation radius is able to follow the shrinking binary inward. But
eventually the gravitational radiation timescale becomes shorter than
the viscous timescale in the disk, leading to a 
merged black hole surrounded
by a hollow disk of gas. 
We show that the subsequent
viscous evolution of the hollow, radiation pressure-dominated
disk will create an $\sim 10^{43.5}(M/10^6M_\odot)\mbox{ergs s}^{-1}$ 
X-ray source on a timescale 
$\sim  7(1+z)(M/10^6M_\odot)^{1.32}\,\mbox{yr}$.
This justifies follow-up monitoring of gravitational wave events with
next-generation X-ray observatories.  Analysis of the detailed light
curve of these afterglows will yield new insights into the subtle
physics of accretion onto massive black holes.


\end{abstract}

\section{Introduction}
\label{sec:intro}

Evidence is mounting that galactic spheroids contain massive black holes (MBHs) in their nuclei.  When two galaxies merge, their MBHs form a binary in the nucleus of the new galaxy \citep{Begelman:80}.  The binary interacts with its stellar and gaseous environment and with other MBHs that collect in the same nucleus in multiple mergers.  These interactions torque the binary and extract its orbital angular momentum and energy.  The binary can thus be rendered so compact that gravitational radiation (GR) carries away the remaining orbital energy, inducing coalescence.  For binaries with masses $M=M_1+M_2\lesssim10^7 M_\odot$, the gravitational slingshot ejection of stars is sufficient to guarantee coalescence in a Hubble time \citep{Milosavljevic:03}.  GR emitted by these binaries shortly before and during coalescence will be detected by the space-based gravitational wave detector {\it Laser Interferometer Space Antenna} ({\it LISA}).\footnote{http://lisa.jpl.nasa.gov.} The coalescence of isolated black holes in general relativity is not accompanied by observable electromagnetic (EM) emission.  We here show that circumbinary gas can lead to a delayed EM afterglow.

Of interest to {\it LISA} are mergers of binaries with masses $M<10^7 M_\odot$ of arbitrary mass ratio $q\leq 1$.  There is abundant evidence for dense gas in galactic nuclei.  Geometrically thin molecular disks have been detected in water maser emission in Seyfert galaxies (e.g., \citealt{Miyoshi:95,Greenhill:03}).  The Galactic nucleus contains a $4\times10^6M_\odot$ MBH surrounded by an $\sim10^4M_\odot$ molecular gas torus (e.g.,~\citealt{Jackson:93}).  Massive accretion disks must be present in quasars and narrow-line Seyfert I nuclei to explain their luminosities. 

In general, the specific angular momentum of inflowing gas exceeds that of the binary, while the gas temperature is below the virial temperature $GM\mu m_{\rm p}/2ak$, where $\mu$ is the mean molecular weight in units of the proton mass $m_{\rm p}$, $a$ is the binary's semimajor axis, and $k$ is the Boltzmann constant.  The sub-virial gas settles into a rotationally supported, geometrically thin circumbinary disk.  If the disk is inclined with respect to the binary's orbital plane, the quadrupole component of the binary's gravitational potential causes differential precession and the warping of the disk.  As in the \citet{Bardeen:75} mechanism, the warp dissipates viscously, resulting in a disk in the binary's orbital plane \citep{Larwood:97,Ivanov:99}.

The disk is truncated at an inner edge 
where gravitational torques and
viscous stresses balance (e.g.,~\citealt{Artymowicz:94}).
Surface density in the hollowed region is much smaller than in the disk (e.g., \citealt{Guenther:04}).
While material does peel off the inner edge and flow across the hole \citep{Artymowicz:96}, the accretion rate for moderate mass ratios ($q\gtrsim 0.01$) is only a fraction ($\lesssim 10\%$) of the rate at which the disk would be accreting without being torqued by the binary (\citealt{Lubow:99}). 
As the binary's semimajor axis decays owing to stellar
or gas dynamical processes (including angular momentum extraction by
the circumbinary disk itself), the disk's inner edge spreads inward
viscously while maintaining an approximately constant ratio of inner edge
radius to semimajor axis, $r_{\rm edge}/a\sim2$.  
In the final stages of in-spiral, however,
the time for decay of the semimajor axis is set by GR and decreases rapidly with $a$ (equation \ref{eq:grtime}).
When this becomes shorter than the (viscous) time
for the inner edge of the disk to spread inward, the binary continues toward
merger while the disk structure remains frozen \citep{Armitage:02}.  
This was also noted by \citet{Liu:03}, who suggested 
that the formation of an inner hole in the disk preceding MBH coalescence causes the interruption of jet activity in double-double radio galaxies.

Vertical support in the central parts of MBH accretion disks is
dominated by radiation pressure, while opacity is dominated by
electron scattering.  The \citet{Shakura:73} $\alpha$-disks in this
regime are thermally and viscously unstable \citep{Lightman:74} if
viscous stresses scale with total pressure.
However, they are stable if the stresses scale
with the gas pressure alone \citep{Sakimoto:81}.  Such a scaling
can occur \citep{Turner:04} if the distance
photons diffuse per orbit is about equal to the scale of magnetic fields
produced by
the magnetorotational instability (MRI; \citealt{Balbus:91})
that facilitates angular momentum transport.  The
radiation then partly decouples from the turbulence driven by the
MRI \citep{Turner:03}. 
These disks are clumpy and porous to the radiation produced
within (\citealt{Begelman:02} and references therein).

We here describe the observable signatures of such circumbinary disks
in MBHs.  In \S~\ref{sec:model}, we present a scenario for the combined
evolution of an MBH binary and its circumbinary accretion disk immediately preceding coalescence.  In \S~\ref{sec:surveys}, we
discuss observable signatures.

\section{A Model For Binary-Disk Evolution}
\label{sec:model}

In the early stages of the GR-driven orbital evolution, the binary
orbit is circularized by GR.  Furthermore, unless $q\ll 1$, 
the timescale \citep{Peters:64}
\beq
\label{eq:grtime}
t_{\rm gr}
\equiv\frac{a}{da/dt}=\frac{5}{64}\frac{c^5a^4}{G^3M^3}\frac{(1+q)^2}{q}.
\eeq 
on which the binary's semimajor axis $a$ decays because the emission of GR greatly {\em exceeds} 
the viscous timescale $t_{\rm visc}=(2/3)r^2/\nu(r)$ for $r\sim a$.
For realistic disks (see below), the effective kinematic viscosity $\nu$ is a weak function of radius, so $t_{\rm visc}\propto r^2$ approximately.  Thus, any gas {\em inside} the binary orbit should long ago have been accreted (for $r\ll a$) or expelled (for $r\sim a$), leaving the MBHs in a gas-free ``donut hole.''\footnote{This is in contrast to the $q\ll 1$ case considered by \citet{Armitage:02}, who assumed that there was still gas accreting onto the black holes.}

Gas outside the binary will attempt to accrete but be prevented by torques from the binary (\S~\ref{sec:intro}).  The inner edge of the disk lies at $r_{\rm edge}=2\lambda a$, where $\lambda>0.5$ is a parameter of order unity.
In this regime, the viscous torque in the disk $3\pi r^2\nu\Sigma\Omega$ is constant with radius (e.g.,~\citealt{Pringle:91}). Here $\Sigma$ is the surface density and $\Omega=(GM/r^3)^{1/2}$ is the local angular velocity.  As the binary shrinks, the inner parts of the circumbinary 
disk pass through a time sequence of nonaccreting, constant torque configurations. In the outer parts of the disk it can occur that $t_{\rm gr}<t_{\rm visc}(r)$, where $r$ is a radius in the disk.  Then material outside $r$ loses viscous causal connection with the inner edge and decouples from the binary torque.

Since $t_{\rm gr}\propto a^4$, while the viscous timescale at $r_{\rm edge}$ approximately scales as $t_{\rm visc}\propto a^2$, the shrinking binary/disk system eventually reaches a state in which $t_{\rm gr}<t_{\rm visc}(r_{\rm edge})$, i.e., the whole disk has decoupled from the rapidly shrinking binary.  The inner parts of the disk begin to evolve as a standard, zero-torque disk. When the inner edge decouples, the radial profile closely resembles the non-accreting disk with a sharp edge.  The inner edge reaches the center in time
$t_{\rm sh}\sim \beta t_{\rm visc}(r_{\rm edge})$,
where
$\beta\equiv \max\{0.1,[d\ln\Sigma/d\ln r(r_{\rm edge})]^{-1}\}$.
The factor 0.1 is appropriate for an infinitely sharp disk
\citep{Lyndenbell:74}.
To estimate $t_{\rm sh}$ and the self-consistent $a$ for which $t_{\rm gr}\sim t_{\rm sh}$, we must determine the internal structure and viscosity of the inner disk at the time of decoupling, to which problem we turn now.  

In the stable
$\alpha$-model, the kinematic viscosity is given by 
$\nu\sim\twothirds\alpha_{\rm gas} P_{\rm gas}/\rho\Omega$, 
where $P_{\rm gas}=\rho kT/\mu m_{\rm p}$ is the gas pressure, 
and $\rho$ is the density.
To prevent confusion, we define 
$\alpha_{\rm gas}$, related to the
usual definition
via $\alpha_{\rm gas}P_{\rm gas}=\alpha
P_{\rm total}$, where $P_{\rm total}=P_{\rm rad}+P_{\rm gas}+P_{\rm
mag}$ is the sum of the radiation, gas, and magnetic pressures. In a simulation of one such accretion disk, \citet{Turner:04}
measures $\alpha=0.0013$ and $P_{\rm rad}/P_{\rm gas}=14$ while
$P_{\rm mag}\ll P_{\rm rad}$, implying that $\alpha_{\rm
gas}\approx0.02$ in his simulation.  We defer modeling the detailed
vertical structure of the disk to a follow-up paper;
relations presented below are approximate.

The rate of local dissipation per unit area of the disk equals ${\cal Q}\sim (9/4)\nu\Sigma\Omega^2$. When we ignore horizontal advection, the flux $F={\cal Q}/2$ must be emitted from each surface. If the surface is emitting as a blackbody, the flux equals $F_{\rm bb}=4\sigma T^4/3\tau$, where $T$ is the midplane temperature, $\tau$ is the total optical depth, and $\sigma$ is the Stefan-Boltzmann constant.  The spectrum differs from that of a blackbody since the photons at different frequencies are thermalized at different depths.  The modified, ``graybody'' spectrum has emergent flux $F_\nu\sim \pi \epsilon_\nu^{1/2}(1+\epsilon_\nu^{1/2})^{-1}B_\nu$, where $B_\nu$ is the Planck function and $\epsilon_\nu=\kappa_{\rm abs,\nu}/(\kappa_{\rm abs,\nu}+\kappa_{\rm es})$ is the ratio of the absorption to the total opacity, and $\kappa_{\rm es}$ is the electron scattering opacity. The quantities $B_\nu$ and $\epsilon_\nu$ are evaluated at the bottom of the thermalization photosphere (TP); we denote the temperature and the density there by $T_\nu$ and $\rho_\nu$, respectively.

The scale height of the photosphere is given by $h_\nu\sim [\gamma
P_{\rm rad}(T_\nu)/\rho_\nu]^{1/2}/\Omega$, where $\gamma\approx4/3$
is the adiabatic index.  We set the absorption opacity equal to the
(mainly bound-free)
opacity $\kappa_{\nu}\approx3\times10^{26}g_{\rm
bf}\rho T^{-7/2}\xi^{-3}(1-e^{-\xi})\textrm{ cm}^2\textrm{ g}^{-1}$, with $\rho$ and $T$ in
cgs, where $\xi\equiv h\nu/kT$, $h$ is the Planck constant, and
$g_{\rm bf}\sim 1$ is the gaunt factor 
($\kappa_\nu$ is the free-free opacity scaled up 
by the ratio of the bound-free to the free-free Rosseland mean opacities
at solar metallicity).
By definition, the effective optical depth
$\sim(\kappa_{{\rm abs},\nu}\kappa_{\rm es})^{1/2}\rho_\nu h_\nu=1$ at the bottom of TP. This is solved for $\rho_\nu$, which
we substitute back in $\kappa_{\rm abs}$ to find that
$\epsilon_\nu^2=1.4\times10^{42}g_{\rm bf}\Omega^2
T_\nu^{-15/2}\xi^{-3}(1-e^{-\xi})$.  To estimate the degree to which
blanketing by the TP modifies the integrated emitted flux, we evaluate
$\epsilon_\nu$ at the Wien frequency
($\xi\approx 2.8$).  We relate $T_\nu$ to the midplane temperature via
$T_\nu=\tau^{-1/4}T$, where $\tau$ is the electron scattering optical
depth between the photosphere and the midplane. This yields 
\bea
\label{eq:flux_grey}
F_{\rm gb}&\equiv& \int_0^\infty F_\nu d\nu 
\sim\frac{4\sigma T^4}{3\tau} 
\frac{\sqrt{\epsilon}}{1+\sqrt{\epsilon}} \sim \frac{9}{8}\nu\Sigma\Omega^2 ,
\eea
where $\epsilon\sim 2.5\times10^{20}\Omega\tau^{15/16}T^{-15/4}$ and the last approximate equality in equation (\ref{eq:flux_grey}) follows from identifying $F_{\rm gb}$ with half of the power generated in the disk. The disk edge has $\epsilon<1$ at decoupling, resulting in a higher midplane temperature than in the blackbody disk.

The optical depth is given by $\tau=\theta\kappa\Sigma$, where $\kappa$ is the opacity (electron scattering $\kappa_{\rm es}\approx0.4\textrm{ cm}^2\textrm{ g}^{-1}$, or Kramer's $\kappa_{\rm abs}\approx 1.6\times10^{24}\rho T^{-7/2}\textrm{ g}^{-1}$) and $\theta\leq1$ is a ``porosity'' correction factor. We introduce $\theta$ to account for the shortened effective optical depth in an inhomogeneous disk in which radiation escapes through low-density domains.  The correction can also be used to approximate disks in which a fraction of the turbulent magnetic energy is dissipated in surface layers \citep{Miller:00} or in the presence of photon-bubble instability \citep{Gammie:98}.  We estimate that $\theta\approx0.2$ in \citet{Turner:04}.  

The surface density is thus far unspecified.  One (admittedly artificial) way to parametrize $\Sigma$ is to fix the accretion rate $\dot M\equiv 3\pi\nu\Sigma$ that the disk would have at its inner edge if it contained a single black hole instead of a binary (the disk is not in viscous equilibrium, and the accretion rate varies with radius).  The accretion rate can be expressed in units of the Eddington rate $\dot M_{\rm edd}\equiv4\pi GMm_{\rm p}/\eta c\sigma_{\rm T}$, where $\eta\sim0.1$ is the radiative efficiency and $\sigma_{\rm T}$ is the Thomson cross section; $\dot M_{\rm edd}$ should not be confused with the {\it local} Eddington limit at the disk edge.  We define the dimensionless parameter $\dot m\equiv\dot M/\dot M_{\rm edd}$.  There is no a priori requirement for $\dot m\leq1$;  in fact, following decoupling the accretion rate will likely be the largest anywhere among black holes of the same mass.  The surface density is thus parametrized via
\beq
\label{eq:sigma_mdot}
\nu\Sigma=\frac{4}{3}\frac{GMm_{\rm p}}{\eta c\sigma_{\rm T}}\dot m .
\eeq
Another way to characterize the surface density is via the total mass of the disk $M_{\rm disk}$ (see Table \ref{tab:gb}), out to the radius where the temperature falls below $10^4\textrm{ K}$ (see below).

To determine the disk properties at the time of decoupling, we
equate $t_{\rm sh}$ with $t_{\rm gr}$, which 
with the expression for the viscosity determines the midplane temperature
near the disk edge in terms of the semimajor axis.
This is then used in equations 
(\ref{eq:flux_grey}) and (\ref{eq:sigma_mdot}) to solve for the
semimajor axis $a$, the edge surface density $\Sigma$, and the edge
midplane temperature $T$ at decoupling. 
Self-consistent solutions 
are in the radiation pressure-dominated and electron scattering-dominated
regime.  The results are
summarized in
Table \ref{tab:gb}.  
We have defined
$\alpha_{\rm gas}=0.1\alpha_{-1}$, $\beta=0.1\beta_{-1}$,
$\eta=0.1\eta_{-1}$, $M=10^6M_6M_\odot$, and
$\theta=0.2\theta_{0.2}$.

We calculate the disk scale height-to-radius ratio $h/r=\Sigma/r\rho$.  For the fiducial choice of parameters $\alpha_{-1}=\beta_{-1}=\eta_{-1}=\lambda=\dot m=M_6=q=\theta_{0.2}=1$, the disk is marginally geometrically thick at the inner edge. We ignore horizontal and vertical advection but do provide an estimate of the advected-to-radiated heat flux ratio ${\cal Q}_{\rm adv}/{\cal Q}_{\rm rad}$.  Ignoring radiative loss, we find the advected flux equals ${\cal Q}_{\rm adv}=(3/4)(4\gamma_p-12\gamma_{\rm T})\nu\Sigma P_{\rm rad}/\rho r^2$, where $\gamma_{\rho,T}\equiv d\ln(\rho,T)/d\ln r$; in our disk, $(4\gamma_p-12\gamma_{\rm T})\sim8$.  In the fiducial disk, horizontal advection is competitive with radiative diffusion, implying that the true disk is geometrically thinner than ours, as in the ``slim-disk'' solutions of \citet{Abramowicz:88}. Just how large could the parameter $\dot m$ be?  The disk must be thin at the edge $(h/r)_{\rm edge}\ll 1$, implying (we ignore the weak dependence on $\alpha$, $M$, and $\mu$)
\bea
\dot m \ll 
1.4 \beta_{-1}^{0.39}\eta_{-1}\lambda^{1.56}[4q/(1+q)^2]^{0.39}\theta_{0.2}^{-0.49} ,
\eea
which allows for $\dot m\gg 1$ when $q\ll 1$.  A trustworthy disk model should also be gravitationally stable, with $Q\equiv\Omega^2/\pi G\rho>1$ throughout. 
We have checked that the disk structures for $M\leq10^6M_\odot$ have $Q>1$ inside the radius where $T$ falls below $10^4\textrm{ K}$ and the disks become susceptible to ionization instability.  We do not extrapolate our disks beyond this radius.

\begin{deluxetable}{cccccccc}
\tablecolumns{8}
\tablecaption{\label{tab:gb}Disk at Decoupling\tablenotemark{a}}
\tablehead{
  \colhead{Variable} &
  \colhead{Factor} &
  \colhead{$\alpha_{-1}$} & 
  \colhead{$\eta_{-1}/\dot m$} &
  \colhead{$\lambda$} &
  \colhead{$M_6$} &
  \colhead{$\beta_{-1},\frac{4q}{(1+q)^2}$} &
  \colhead{$\theta_{0.2}$}
}
\startdata 
$a/(GM/c^2)$ & 117 & -0.34 & 0.24 & 0.70 & 0.08 & 0.42 & -0.08 \\
$T\ (10^6\textrm{ K})$ & 1.7 & 0.19 & -0.86 & -1.95 & -0.28 & -0.49 & 0.30 \\
$t_{\rm sh}\  (\textrm{yr}$) & 9.4 & -1.36 & 0.98 & 2.80 & 1.32 & 1.7,\ 0.7 & -0.34 \\
$h/r$ & 0.46 & 0.76 & -2.43 & -3.80 & -0.12 & -0.95 & 1.19 \\
$P_{\rm rad}/P_{\rm gas}$ & 2600 & 1.67 & -4.25 & -7.35 & -0.04 & -1.84 & 2.17 \\
$\kappa_{\rm es}/\kappa_{\rm abs}$ & 46,000 & 1.76 & -4.68 & -8.32 & -0.18 & -2.08 & 2.32 \\
$Q\ (10^5)$ & 4.5 & 2.12 & -2.41 & -6.61 & -1.44 & -1.65 & 1.53 \\
${\cal Q}_{\rm adv}/{\cal Q}_{\rm rad}$ & 0.44 & 1.52 & -4.86 & -7.60 & -0.24 & -1.90 & 2.38 \\
$\epsilon\ (10^{-3})$ & 1.4 & -0.84 & 2.37 & 4.20 & 0.08 & 1.05 & -0.21 \\
$M_{\rm disk}\ (M_\odot)$ & 96 & -1.17 & -0.88 & -1.15 & 2.04 & 0.21 & -0.04 \\
\enddata
\tablenotetext{a}{Variable in column 1, defined in the text, equals the factor in column 2 times the product of column head parameters raised to the exponents indicated in columns 3--8, where 
we assume $\mu=0.6$. All quantities except for $M_{\rm disk}$ are evaluated at the edge of the disk.}

\end{deluxetable}

\section{Detection of the Electromagnetic Afterglow}
\label{sec:surveys}

\begin{figure}
\plotone{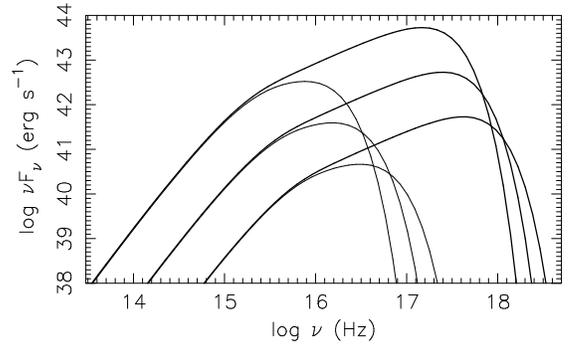}
\caption{\label{fig:spectra} Representative thermal disk spectra before the coalescence ({\it thin lines}) and after ({\it thick lines}). The spectra are modified blackbody spectra of Eddington-limited ($\dot m=1$) accretion disks around black holes of mass $M=(10^4,10^5,10^6)M_\odot$ ({\it from right to left}) and dimensionless spin parameter $a_*=0.9$ (we ignored Doppler broadening). Only the emission from rings in the disk with $r<1000GM/c^2$ was taken into account.  These crude spectra are compatible with detailed models of \citet{Hubeny:01}.  We do not show thermal emission from any accretion disks around individual black holes that may be fed by the gas flow crossing the hollowed region.
}
\end{figure}

The pre-coalescent circumbinary disk is expected to be luminous in IR with a negligible X-ray counterpart (Figure \ref{fig:spectra}).  Unfortunately, this IR source may be confused with the light of the host galaxy.  \citet{Barth:04} recently studied the dwarf galaxy POX 52, which appears to contain a $1.6\times10^5M_\odot$ black hole accreting at about the Eddington rate (\citealt{Greene:04} list other similar candidates).  The optical luminosity of the POX 52 black hole is about the same as that of the galaxy, so it seems unlikely that the MBH binary can be identified electromagnetically before the merger.

The black holes merge $t_{\rm gr}/4$ after decoupling, where $t_{\rm gr}$ is evaluated at decoupling. The gravitational wave losses during merger immediately perturb the potential in which the surrounding gas orbits, giving a (weak) prompt EM signature. Subsequently, the inner edge of the accretion disk migrates inward on timescale $\sim t_{\rm sh}\sim t_{\rm gr}$, arriving at the merged black hole $\sim\threequarters t_{\rm sh}$ after the merger.  Its arrival is accompanied by rapid accretion and the activation of an X-ray active galactic nucleus (AGN).  
We consider these two distinct types of EM signatures in turn.

The few percent reduction in the total black hole mass due to gravitational wave losses excites a weak axisymmetric wave in the disk. More interestingly,
the coalescence may be accompanied by radiation recoil (e.g.,~\citealt{Merritt:04}, and references therein). The velocity of the recoiled black hole, $v_{\rm rec}\lesssim 300\textrm{ km s}^{-1}$, is much smaller than the orbital velocity of the disk at the inner edge, $\sim 2\times10^4\textrm{ km s}^{-1}$. Disk material at radius $r$ remains entirely bound to the black hole if its orbital velocity prior to the recoil $(GM/r)^{1/2}$ is larger than $\sim 2.4v_{\rm rec}$, which is true for the inner disk. The recoil drives waves and warps in the outer disk. If these result in shocks or obscuration, they could result in an observable EM signature.    

\begin{figure}
\plotone{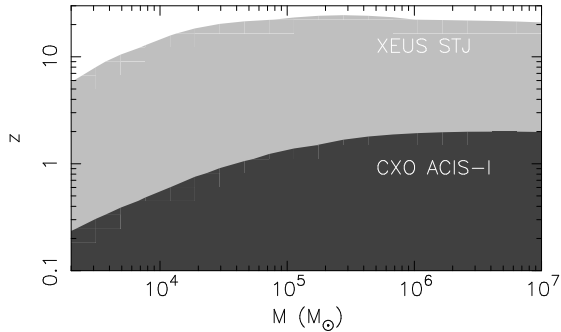}
\caption{\label{fig:redshift} Redshifts at which emission from an Eddington-accreting ($\dot m=1$) black hole of mass $M$ can be detected with $10^5\textrm{ s}$ exposure using the instrument ACIS-I on {\it Chandra} ({\it dark gray}) and the Hafnium superconducting tunneling junction (STJ) detector on XEUS before mission upgrade (light-grey).  We assume that 10 counts constitute detection and ignore absorption and confusion. The disk spectra were thermal modified blackbody spectra for a nearly maximally spinning black hole, $a_*=0.9$. We assume the standard $\Lambda$CDM cosmology with $H_0=70\textrm{ km s}^{-1}\textrm{ Mpc}^{-1}$, $\Omega_{\rm m}=0.3$, and $\Omega_\Lambda=0.7$.
}
\end{figure}

More certain is the afterglow due to the viscous migration of the inner edge of the disk and consequent turn-on of X-ray emission from a rapidly accreting AGN. {\it LISA} will detect with a high signal-to-noise ratio the coalescence of black holes with masses $(10^{5}-10^{7})(1+z)^{-1}M_\odot$ at redshifts $z\lesssim20$. Can the EM afterglow also be detected? 
From Table \ref{tab:gb}, we estimate the observed interval $\Delta t\sim \threequarters(1+z)t_{\rm sh}$ between the two signals.  For example, for the merger of two $10^5M_\odot$ black holes at $z=5$, assuming $\alpha_{-1}=\beta_{-1}=\eta_{-1}=\dot m=\theta_{0.2}=1$, we find $\Delta t\sim2\textrm{ yr}$ for a graybody disk truncated at $r_{\rm edge}=2a$.  A strong recoil could shorten this.
Most of the luminosity of the postcoalescent accretion disk around a $(10^4-10^6)M_\odot$ black hole is emitted at rest-frame energies $h\nu\sim(0.5-5)\textrm{ keV}$ (see Figure \ref{fig:spectra}), which fall within the sensitivity windows of future X-ray detectors such as {\it XEUS}\footnote{http://www.rssd.esa.int/index.php?project=XEUS.} and {\it Generation-X}.\footnote{http://generation-x.gsfc.nasa.gov.}  {\it XEUS} could see this emission at $z\sim20$ for $\gtrsim10^5M_\odot$ black holes.  For smaller black holes, the maximum redshift for detection decreases with the decreasing mass (see Figure \ref{fig:redshift}). 

At $z\lesssim 20$, the {\it James Webb Space Telescope}\footnote{http://www.stsci.edu/jwst.} is sensitive to rest-frame mid-UV to near-IR, which
come from the disk radii exterior to the inner edge at decoupling (Figure \ref{fig:spectra}).  Therefore, the post-coalescence IR--optical flux will not be very different from the pre-coalescence flux.  An exception would be if the merged black hole were enshrouded by a large column depth of gas and dust. Then, the UV--X-ray emission would be reprocessed to longer wavelengths. Such a shrouded merger could be identified by a sudden increase in IR luminosity.

We therefore propose that locations with detected GR signals from MBH coalescence be monitored in the $0.1-10\textrm{ keV}$ band at high angular resolution. The angular resolution of {\it LISA} ranges from several arcminutes to several degrees depending on the black hole mass, mass ratio, and redshift \citep{Cutler:98,Hughes:02}. Therefore, multiple exposures of an X-ray telescope may be necessary for some sources.  A possibly confusing source of X-ray flares are tidal disruptions of mainsequence stars \citep{Rees:90,Cannizzo:90,Komossa:04}.  Detection of the afterglow of MBH coalescence will help pinpoint the location and hence redshift of GR sources. The detailed light curve of the afterglow  will probe the structure of the accretion disk as it moves toward the black hole, and will also shed light on the cosmological assembly of MBHs.

\acknowledgements

We thank Aaron Barth, Tsvi Piran, and Tom Prince for valuable discussions. M.~M.~was supported at Caltech by a postdoctoral fellowship from the Sherman Fairchild Foundation. E.~S.~P. was supported in part by NASA ATP grants NAG5-10707 and NNG04GK98G.


\begin{thebibliography}{}

\bibitem[Abramowicz et al.(1988)]{Abramowicz:88}
Abramowicz, M.~A., Czerny, B., Lasota, J.~P., \& Szuszkiewicz, E.\ 1988, \apj, 332, 646

\bibitem[Armitage \& Natarajan(2002)]{Armitage:02} Armitage, 
P.~J.~\& Natarajan, P.\ 2002, \apjl, 567, L9 

\bibitem[Artymowicz \& Lubow(1994)]{Artymowicz:94} Artymowicz, P.~\& 
Lubow, S.~H.\ 1994, \apj, 421, 651 

\bibitem[Artymowicz \& Lubow(1996)]{Artymowicz:96} Artymowicz, P.~\& 
Lubow, S.~H.\ 1996, \apjl, 467, L77 

\bibitem[Balbus \& Hawley(1991)]{Balbus:91} Balbus, S.~A.~\& 
Hawley, J.~F.\ 1991, \apj, 376, 214 

\bibitem[Bardeen \& Petterson(1975)]{Bardeen:75} Bardeen, J.~M.~\& 
Petterson, J.~A.\ 1975, \apjl, 195, L65 

\bibitem[Barth et al.(2004)]{Barth:04} Barth, 
A.~J., Ho, L.~C., Rutledge, R.~E., \& Sargent, W.~L.~W.\ 2004, \apj, 607, 
90 

\bibitem[Begelman(2002)]{Begelman:02} Begelman, M.~C.\ 2002, \apjl, 568, L97

\bibitem[Begelman, Blandford, \& Rees(1980)]{Begelman:80} Begelman, 
M.~C., Blandford, R.~D., \& Rees, M.~J.\ 1980, \nat, 287, 307 

\bibitem[Cannizzo, Lee, \& Goodman(1990)]{Cannizzo:90}
Cannizzo, J.~K., Lee, H.~M., \& Goodman, J.\ 1990, \apj, 351, 38

\bibitem[Cutler(1998)]{Cutler:98} Cutler, C.\ 1998, \prd, 57, 7089

\bibitem[Gammie(1998)]{Gammie:98} Gammie, C.~F.\ 1998, \mnras, 297, 929

\bibitem[Greene \& Ho(2004)]{Greene:04} Greene, J.~E.~\& Ho, 
L.~C.\ 2004, \apj, 610, 722 

\bibitem[Greenhill et al.(2003)]{Greenhill:03} Greenhill, L.~J., et 
al.\ 2003, \apj, 590, 162 

\bibitem[G{\" u}nther, Sch{\" a}fer, \& Kley(2004)]{Guenther:04}
G{\" u}nther, R., Sch{\" a}fer, C., \& Kley, W.\ 2004, \aap, 423, 559

\bibitem[Hubeny et al.(2001)]{Hubeny:01} 
Hubeny, T., Blaes, O., Krolik, J.~H., \& Agol, E.\ 2001, \apj, 559, 680

\bibitem[Hughes(2002)]{Hughes:02} Hughes, S.~A.\ 2002, \mnras, 331, 805

\bibitem[Ivanov, Papaloizou, \& Polnarev(1999)]{Ivanov:99} 
Ivanov, P.~B., Papaloizou, J.~C.~B., \& Polnarev, A.~G.\ 1999, \mnras, 307, 79 

\bibitem[Jackson et al.(1993)]{Jackson:93} Jackson, J.~M., Geis, 
N., Genzel, R., Harris, A.~I., Madden, S., Poglitsch, A., Stacey, G.~J., \& 
Townes, C.~H.\ 1993, \apj, 402, 173 

\bibitem[Komossa et al.(2004)]{Komossa:04} Komossa, S., Halpern, J., Schartel, N., Hasinger, G., Santos-Lleo, M., \& Predehl, P.\ 2004, \apjl, 603, L17

\bibitem[Larwood \& Papaloizou(1997)]{Larwood:97}
Larwood, J.~D.~\& Papaloizou, J.~C.~B.\ 1997, \mnras, 285, 288

\bibitem[Lightman \& Eardley(1974)]{Lightman:74} Lightman, A.~P.~\& 
Eardley, D.~M.\ 1974, \apjl, 187, L1 

\bibitem[Liu, Wu, \& Cao(2003)]{Liu:03} Liu, F.~K., Wu, X., \& Cao, S.~L.\ 2003, \mnras, 340, 411

\bibitem[Lubow, Seibert, \& Artymowicz(1999)]{Lubow:99}
Lubow, S.~H., Seibert, M., \& Artymowicz, P.\ 1999, \apj, 526, 1001

\bibitem[Lynden-Bell \& Pringle(1974)]{Lyndenbell:74}
Lynden-Bell, D. \&\ Pringle, J.E.\ 1974, \mnras, 168, 603

\bibitem[Merritt et al.(2004)]{Merritt:04} Merritt, D., Milosavljevi{\' c}, M., Favata, M.,
Hughes, S.~A., \& Holz, D.~E.\ 2004, \apjl, 607, L9

\bibitem[Miller \& Stone(2000)]{Miller:00} Miller, K.~A.~\& Stone, J.~M.\ 200, \apj, 534, 398
 
\bibitem[Milosavljevi{\' c} \& Merritt(2003)]{Milosavljevic:03}
Milosavljevi\'c, M.~\& Merritt, D.\ 2003, \apj, 596, 860

\bibitem[Miyoshi et al.(1995)]{Miyoshi:95} Miyoshi, M., Moran, J., 
Herrnstein, J., Greenhill, L., Nakai, N., Diamond, P., \& Inoue, M.\ 1995, 
\nat, 373, 127 

\bibitem[Peters(1964)]{Peters:64} Peters, P.~C.\ 1964, Physical 
Review , 136, 1224

\bibitem[Pringle(1991)]{Pringle:91} Pringle, J.~E.\ 1991, \mnras, 
248, 754 

\bibitem[Rees(1990)]{Rees:90} Rees, M.~J.\ 1990, Science, 247, 817

\bibitem[Sakimoto \& Coroniti(1981)]{Sakimoto:81} Sakimoto, 
P.~J.~\& Coroniti, F.~V.\ 1981, \apj, 247, 19 

\bibitem[Shakura \& Sunyaev(1973)]{Shakura:73} Shakura, N.~I.~\& 
Sunyaev, R.~A.\ 1973, \aap, 24, 337 

\bibitem[Turner(2004)]{Turner:04} Turner, N.~J.\ 2004, \apjl, 
605, L45 

\bibitem[Turner et al.~(2003)]{Turner:03} Turner, 
N.~J., Stone, J.~M., Krolik, J.~H., \& Sano, T.\ 2003, \apj, 593, 992 

\end{thebibliography}
\end{document}